\documentclass[apj,graphicx]{emulateapj-rtx4}
\usepackage{xcolor}

\newcommand{\src}{Swift J1753.5--0127} 

\newcommand{\swift}{\textit{Swift}}

\newcommand{\Msun}{\mathrm{M}_{\odot}}

\begin{document}

\shorttitle{Mini-outburst of Swift J1753.5--0127}
\shortauthors{Zhang et al.}

\title{Bright mini-outburst ends the 12$-$year long activity of the 
black hole candidate Swift J1753.5$-$0127}

\author{G.-B.~Zhang\altaffilmark{1,2,3,4}}
\author{F.~Bernardini\altaffilmark{5,6,4}}
\author{D.M.~Russell\altaffilmark{4}}
\author{J.D.~Gelfand\altaffilmark{4,7}}
\author{J.-P.~Lasota\altaffilmark{8,9}}
\author{A.~Al Qasim\altaffilmark{4,10}}
\author{A.~AlMannaei\altaffilmark{4,10}}
\author{K.~I.~I.~Koljonen\altaffilmark{11, 12}}
\author{A.W.~Shaw\altaffilmark{13}}
\author{F.~Lewis\altaffilmark{14,15}}
\author{J.A.~Tomsick\altaffilmark{16}}
\author{R.M.~Plotkin\altaffilmark{17}}
\author{J.C.A.~Miller-Jones\altaffilmark{17}}
\author{D.~Maitra\altaffilmark{18}}
\author{J.~Homan\altaffilmark{19,20}}
\author{P.A.~Charles\altaffilmark{21}}
\author{P.~Kobel\altaffilmark{22}}
\author{D.~Perez\altaffilmark{22}}
\author{R.~Doran\altaffilmark{23}}

\email{zhangguobao@ynao.ac.cn}

\affil{\altaffilmark{1}{Yunnan Observatories, Chinese Academy of Sciences (CAS), Kunming 650216, P.R. China; Email: zhangguobao@ynao.ac.cn}}
\affil{\altaffilmark{2}{Key Laboratory for the Structure and Evolution of Celestial Objects, CAS, Kunming 650216, P.R. China}}
\affil{\altaffilmark{3}{Center for Astronomical Mega-Science, CAS, Beijing, 100012, P. R. China}}
\affil{\altaffilmark{4}{New York University Abu Dhabi, P.O. Box 129188, Abu Dhabi, United Arab Emirates}}
\affil{\altaffilmark{5}{INAF $-$ Osservatorio Astronomico di Roma, via Frascati 33, I-00040 Monteporzio Catone, Roma, Italy}}
\affil{\altaffilmark{6}{INAF $-$ Osservatorio Astronomico di Capodimonte, Salita Moiariello 16, I-80131 Napoli, Italy}}
\affil{\altaffilmark{7}{Center for Cosmology and Particle Physics, New York University, Meyer Hall of Physics, 4 Washington Place, New York, NY 10003, USA}}
\affil{\altaffilmark{8}{Institut d'Astrophysique de Paris, CNRS et Sorbonne Universit\'es, UPMC Paris~06,  UMR 7095, 98bis Bd Arago, 75014 Paris, France}}
\affil{\altaffilmark{9}{Nicolaus Copernicus Astronomical Center, Bartycka 18, 00-716 Warsaw, Poland}}
\affil{\altaffilmark{10}{Mullard Space Science Laboratory, University College London, Holmbury St. Mary, Dorking, Surrey RH5 6NT, UK}}
\affil{\altaffilmark{11}{Finnish Centre for Astronomy with ESO (FINCA), University of Turku, V\"ais\"al\"antie 20, 21500 Piikki\"o, Finland} } 
\affil{\altaffilmark{12}{Aalto University Mets\"ahovi Radio Observatory, PO Box 13000, FI-00076 Aalto, Finland}}
\affil{\altaffilmark{13}{Department of Physics, University of Alberta, 4-181 CCIS, Edmonton, AB T6G 2E1, Canada}}
\affil{\altaffilmark{14}{Faulkes Telescope Project, School of Physics, and Astronomy, Cardiff University, The Parade, Cardiff, CF24 3AA, Wales, UK}}
\affil{\altaffilmark{15}{Astrophysics Research Institute, Liverpool John Moores University, 146 Brownlow Hill, Liverpool L3 5RF, UK}}
\affil{\altaffilmark{16}{Space Sciences Laboratory, 7 Gauss Way, University of California, Berkeley, CA 94720-7450, USA}}
\affil{\altaffilmark{17}{International Centre for Radio Astronomy Research-Curtin University, GPO Box U1987, Perth, WA 6845, Australia}}
\affil{\altaffilmark{18}{Department of Physics and Astronomy, Wheaton College, Norton, MA 02766, USA}}
\affil{\altaffilmark{19}{Eureka Scientific, Inc., 2452 Delmer Street, Oakland, CA 94602, USA}}
\affil{\altaffilmark{20}{SRON, Netherlands Institute for Space Research, Sorbonnelaan 2, 3584 CA Utrecht, The Netherlands}}
\affil{\altaffilmark{21}{Department of Physics \& Astronomy, University of Southampton, Southampton, SO17 1BJ, UK}}
\affil{\altaffilmark{22}{Gymnase du Bugnon-S\'evelin, Avenue de S\'evelin 44, 1004 Lausanne, Switzerland}}
\affil{\altaffilmark{23}{NUCLIO - N\'ucleo Interactivo de Astronomia, Largo dos Top\'azios, 48, 3 Frt, PT2785--817 S. D. Rana, Portugal}}

\newcommand*\KK{ }
\newcommand*\FB{ }
\newcommand*\GB{ }

\begin{abstract}

We present optical, UV and X-ray monitoring of the short orbital period 
black hole X-ray binary candidate \src, focusing on the final stages of its 12$-$year 
long outburst that started in 2005. 
From September 2016 onward, the source started to fade and within three months, 
the optical flux almost reached the quiescent level. Soon after that, 
{ using a new  proposed rebrightening classification method }
we recorded { a mini-outburst and a reflare} {\KK in the optical light curves,} peaking in February (V$\rm\sim$17.0) 
and May (V$\rm\sim$17.9) 2017, { respectively.}  Remarkably, the { mini-outburst} 
has a peak flux consistent with the extrapolation of the slow decay before the fading 
phase preceding it. The { following reflare} {\KK was} fainter and shorter. We found from optical colors 
that the temperature of the outer disk was $\sim 11$,000 K when the source started to 
fade rapidly. According to the disk instability model, this is 
close to the critical temperature when a cooling wave is expected to form 
in the disk, shutting down the outburst.  The optical color could be a 
useful tool to predict decay rates in some X-ray transients. 
{ We notice that all X-ray binaries that {\KK show} mini-outbursts following 
{\KK a main} outburst are short orbital period systems ($<$ 7 h).}
In analogy with another class of short 
period binaries showing similar mini-outbursts, the cataclysmic variables of the RZ
LMi type, we suggest mini-outbursts could occur if there is a hot 
inner disk at the end of the outburst decay.

\end{abstract}


\section{Introduction}
\label{introduction}

Black hole X-ray transients (BHXTs) are binary systems consisting 
of a stellar-mass black hole (BH) primary accreting matter from a non-collapsed donor 
secondary. The vast majority spend most of their time in quiescence, 
where the X-ray luminosity is low. Episodically, during outburst, the X-ray 
flux increases by orders of magnitude, approaching in some cases the Eddington luminosity 
limit ($L_{\rm Edd}$). BHXTs exhibit a number of  different spectral states 
during outbursts \citep[e.g. low-hard, high-soft; ][]{Belloni16}, where the accretion 
properties change. 

BHXT outbursts are thought to be triggered by a thermal-viscous instability 
in the accretion disk. In the disk instability model
\citep[DIM, for a review see][]{Lasota01}, as matter accumulates in the disk 
during quiescence, the surface density increases until a threshold is exceeded 	
at a certain radius in the disk. Hydrogen becomes ionized, the disk heats up
quickly and the source goes into outburst. 

The DIM can broadly explain the 
quiescent to outburst cycle of BHXTs. 
However, some observed properties cannot be well reproduced by {\FB it}. 
Most BHXTs directly fade into quiescence with an exponential 
decay  after the outburst peak, whereas some BHXTs \citep[e.g. GRO J0422$+$32, 
GRS 1009$-$45, MAXI J1659--152, GRS 1739--278;][]{Chen97, Homan13, Yan17}, 
show several rebrightenings after the outburst {\FB peak}. {\FB Some of} these rebrightenings have been defined as 
mini-outbursts by \cite{Chen97} because {\FB they happen soon after the flux reaches close to the quiescence level, and they are} of  smaller amplitude and shorter duration compared to the 
normal outburst \citep[see e.g.][ hereafter D01]{Dubus01}. 
{ A sudden reactivation soon after the end of an outburst decay has 
been observed also in neutron star (NS) systems \citep[e.g.][]{Hartman11,Patruno16}.}
Mini-outbursts 
are of much greater amplitude than the more common ``reflares'' that typically 
occur during the fade of the main outburst before reaching quiescence \citep{Lasota01}. 
Reflares are theoretically expected to arise from a sequence of heating and cooling 
front reflections in the disk according to the DIM, but could alternatively be caused 
by: 
\begin{enumerate}
\item X-rays heating the companion, {\KK increasing the} mass accretion {\KK  rate}
\citep[outburst `echos'; 
see][and references therein]{Lasota01,Dubus01,Kalemci14},
\item synchrotron emission from the reactivation 
of the jet in the hard state decay \citep[e.g.][]{Kalemci13}, 
\item in the case of neutron star accretors, the activation/deactivation of the propeller 
effect {\KK reducing/increasing the mass accretion rate} due to the rapidly rotating neutron star 
magnetosphere \citep{Hartman11,Patruno16}.
\end{enumerate}

Even rarer are `multi-peak' outbursts, where a second peak (or {\KK several peaks}) following the main outburst 
{\KK reach a} similar flux {\KK level} to the first peak and may last longer \citep[e.g. the BH systems 
GRO J1655--40; XTE J1118+480 and the neutron star system IGR J00291+5934;][]{Chen97,Chaty03,Lewis10}. 
The properties of mini-outbursts are not well explained by the DIM, but require specific 
conditions of the accretion flow such as a hot inner disk residing near the end of the first outburst. 
This has been adopted to explain how the DIM can {\FB reproduce} the mini-outbursts of some 
{\KK dwarf novae \citep[DN; e.g.][]{Hameury00},}
but to date has not successfully been applied to mini-outbursts of BHXTs.

\src\ (hereafter J1753) was discovered by the \swift\ Burst Alert Telescope 
({\it BAT}) on 2005 June 30 \citep{ATel546} and showed a typical 
fast rise, exponential-decay (FRED) light curve. However, after decaying 
from the outburst peak, instead of returning to quiescence, J1753 
remained active, with {\it BAT} fluxes varying between 0.001 and 0.03 count/s (15$-$50 keV).
The X-ray spectral and timing properties during outburst indicate that the source is a BH 
X-ray binary that remained in the low-hard state  
\citep[e.g.][]{CadolleBel07, Zhang07} for the vast majority of the time \citep{Shaw16a}.

%($P_{\it orb}$)
The orbital period {\KK of J1753} is constrained to be 3.24 hours \citep{Zurita08}, 
with an alternative claim of 2.85 hours proposed by \cite{Neustroev14} which has been 
challenged by \cite{Shaw16a}. The source distance is poorly constrained, $\sim$2.5$-$8 kpc, 
and the inclination $i$ is constrained by model-dependent methods to be $\sim 55^{\circ}$
\citep[e.g.][]{Zurita08,Froning14,Plotkin17,Gandhi18}. Time-resolved optical spectroscopy suggests {\KK that}
the primary mass is greater than $7.4\Msun$ \citep{Shaw16a}.
{\KK During the outburst} the radio flux of J1753 \citep{CadolleBel07, Soleri10},  
which is consistent with optically thick synchrotron {\KK emission} likely originating in a compact jet 
\citep[e.g.][]{Tomsick15}, is typically lower than that of BHXTs following the 
standard track in the  radio vs. X-ray diagram \citep{Gallo12,Rushton16}, 
making J1753 one of the radio-faint BHXTs. However, recent observations show that 
the source was close to the standard track during the rise and decay of the 
{\KK rebrightenings discussed below} \citep{Plotkin17}.

After being active for $\sim$11 years, J1753 became faint at X-ray, UV, optical, 
and radio wavelengths, almost reaching the {\KK previous measured  optical quiescence level} 
\citep[V$\sim21$ mag,][]{CadolleBel07} in November 2016 \citep{ATel9708,ATel9735,ATel9739,ATel9765}.
Unexpectedly, within $\sim3$ months J1753 was bright again at 
all wavelengths \citep[e.g.][]{ATel10075, ATel10097, ATel10110}.

Here, we present seven years of multi-band optical monitoring of J1753 with the Faulkes 
and Las Cumbres Observatory (LCO) telescopes, complemented with archival optical data, 
and UV and soft X-ray coverage provided by \swift. In particular, we concentrate on the latest 
stage of the outburst (2016 June$-$2017 June), when the source faded and exhibited {\KK rebrightening episodes.}

\section{Rebrightenings classification}
\label{rebrightenings}

\begin{figure*}
\begin{center}
\includegraphics[width=7.1in,angle=0]{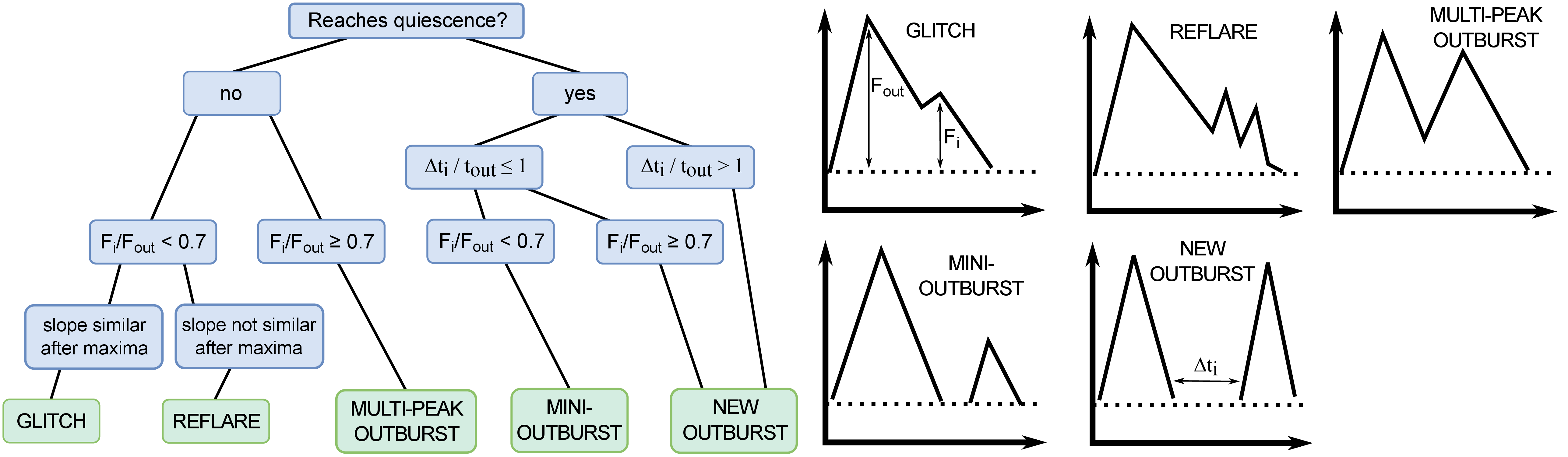}
\caption{{\KK Flowchart of the rebrightening classification scheme (left) with illustrations of the different 
classes of rebrightenings (right).}}
\label{rebrightening-figure}
\end{center}
\end{figure*}

{ 

{\KK Previously,} some of the rebrightening phenomena (e.g., reflares, {\KK  glitches}, mini-outbursts) have been 
classified qualitatively in \cite{Chen97}. 
{\KK Since then}, more rebrightening phenomena in X-ray transients have been 
discovered and labelled (e.g., multi-peak outbursts, secondary outbursts). %and some of rebrightening phenomena were labelled inconsistently with that in the previous literatures. 
{\KK However, a systematic classification scheme has not been developed for the rebrightenings, which has caused several inconsistencies in the literature \citep[e.g. V404 Cyg][]{MunozDarias17}}

In order to classify the rebrightening phenomena quantitatively, we introduce here a new {\FB observation-based} classification scheme, 
{\KK which first asks whether or not the source has reached quiescence {\KK preceding each rebrightening episode.}}
We consider a source to reach quiescence if either (a) 
{\KK the optical magnitude is} within 0.5 mag of the quiescent level (or the X-ray flux is {\KK within} a factor of 2 {\KK of the} X-ray {\KK quiescent level}), 
or (b) if the extrapolation of the {\KK outburst} decay {\KK rate} reaches the quiescent level before the rebrightening is detected. 
{\FB Then we define $F_{i}/F_{out}$ as the flux ratio between each rebrightening peak flux and the outburst peak flux, $\Delta t_{\rm i}$ as the time separating the start of the quiescent period from the start of each rebrightening, and $t_{\rm out}$ as the duration of 
the main outburst.}

{\FB In practice, for a rebrightening event  (following Fig. \ref{rebrightening-figure}):}

\begin{enumerate}

\item If the source flux does not reach {\KK quiescence} before the rebrightening, it can be classified as:

\begin{enumerate}
\item { A glitch}, {\KK if} 
{\FB $F_{i}/F_{out} < 0.7$ and the light curve slope in the decay of the rebrightening is similar to 
the light curve slope during the main outburst decay.} 
%{\FB \citep[see Fig. 9 and 20 and in][]{Chen97}}. 

\item { A reflare}, {\KK if} 
$F_{i}/F_{out} < 0.7$, {\FB but the slopes of the light curve of the outburst decay and that 
following the rebrightening are different.}

\item { A multi-peak outburst}, {\KK if} 
{\FB $F_{i}/F_{out} \geq 0.7$.} 

\end{enumerate}

Since multiple rebrightening events can occur, the above can be used to classify each subsequent 
rebrightening until the flux reaches quiescence.

\item If the flux reaches quiescence before a rebrightening, {\FB it} can be classified as:

\begin{enumerate}

\item { A mini-outburst}, {\KK if} 
{\FB $\Delta t_{\rm i} / t_{\rm out} \leq 1$ and $F_{i}/F_{out} < 0.7$.}

\item { A new outburst}, {\KK if} 
$\Delta t_{\rm i} / t_{\rm out} > 1$
or 
$\Delta t_{\rm i} / t_{\rm out} \leq 1$ and 
$F_{i}/F_{out} \geq 0.7$.

\end{enumerate}

{\FB In the case of a mini-outburst: until the flux reaches quiescence, any following rebrightenings have to be compared with the mini-outburst properties since they are part of the mini-outburst; once the flux reaches quiescence again, the following rebrightening has to be compared with the main outburst.}

\end{enumerate}

{\FB This method provides a unique classification for most of the rebrightenings observed in BH and NS transients.} {\KK However,}
some {\FB caveats} have to {\KK be taken} into account: 
{\FB (1) The classification depends on the energy band (e.g. for the same source, simultaneous optical and X-ray light curves could lead to different classifications).}
{\KK (2) If the source is recurrent (e.g. GX 339--4, for which $\Delta t_{\rm i} / t_{\rm out} \leq 1$ is typically true for a large fraction of its outbursts), all rebrightenings should be categorized as new outbursts.}
(3) If there are {\FB observational} gaps {\FB in the light curves} with {\FB duration} comparable or longer than the 
outburst {\KK length} 
only 
{\FB a potential classification can be given}. 
(4) {\FB We chose $F_{i}/F_{out}=0.7$ to separate strong and weak rebrightenings, and $\Delta t_{\rm i} / t_{\rm out}=1$ to separate mini-outbursts from new outbursts, so that known rebrightenings fall under the same classification as the definitions in \cite{Chen97}. These values can be further tuned in the future, when more data will become available.} 
}

\section{Observations and data reduction}
\label{data}

\subsection{Optical photometry} 
\label{optical data}

\begin{figure*}
    \centering
        \includegraphics[width=7.1in, angle=0]{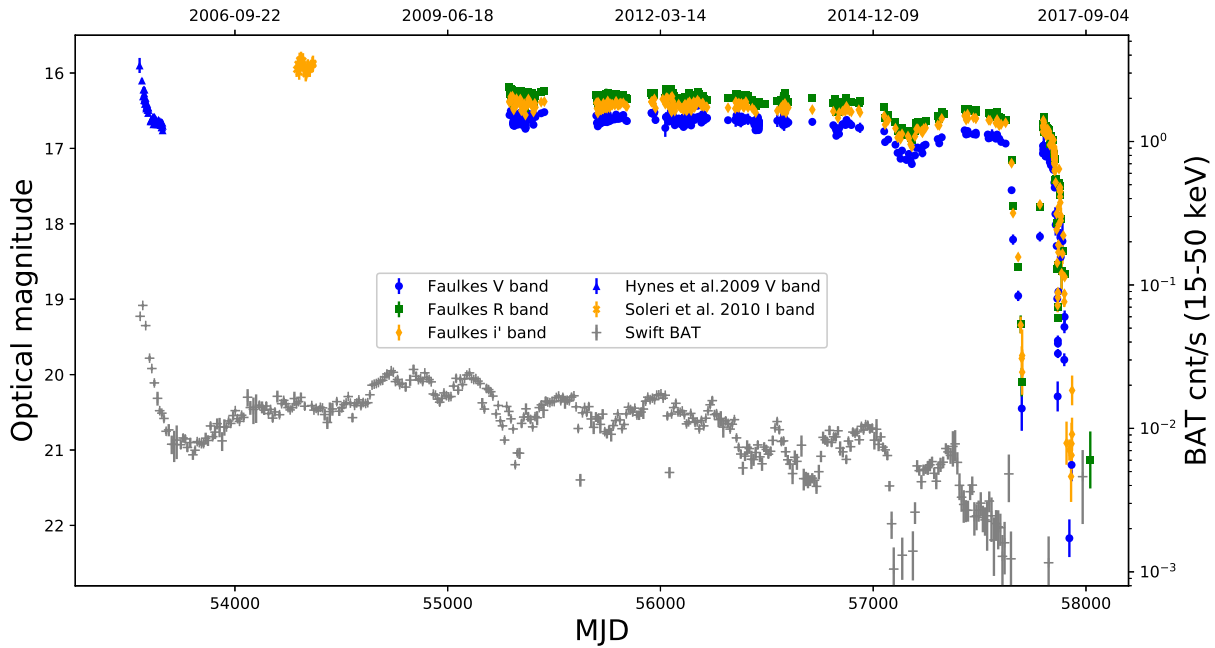}
    \caption{The optical (SMARTS data from \citealt{Hynes09} and \citealt{Soleri10}, and 
    our Faulkes/LCO data) and 15$-$50 keV X-ray (5-d binned) \swift-BAT light curve of J1753.
    Note the { unusual} low-luminosity soft state around MJD 57100, the initial decay toward quiescence 
    around MJD 57650, and the {\KK rebrightening episodes} at MJD $\sim$ 57800--57900.
    }
    \label{fig_lightcurve_x_uv}
\end{figure*}

We observed J1753 with the 2-m robotic Faulkes Telescopes North 
(located at Haleakala on Maui, USA) and South (at Siding Spring, 
Australia)  \citep[e.g.][]{Lewis18}, and the 1-m Las Cumbres Observatory (LCO) telescopes \citep{Brown13}.
These observations are part of an ongoing monitoring campaign of $\sim 40$ 
low-mass X-ray binaries \citep{Lewis08}.
Data were taken in the $V$, $R$, and $i\arcmin$-band filters from 2010 April, 
and were reduced using the LCO automatic pipeline. Photometry was performed 
using \textsc{PHOT} in \textsc{IRAF}. Photometric calibration was achieved 
using stars with magnitude errors of $<0.05$ in the Pan-STARRS1 and APASS 
catalogues \citep{PanStarrs,APASS}.  The 2010--2013 January data were published 
in \cite{Shaw13} using flux calibration from stars listed in \cite{Zurita08}. 
We also {\KK use} archival $I$ and $V$-band data of J1753 published in \cite{Hynes09} 
and \cite{Soleri10}. { The observation logs are provided in Table \ref{tab:obs} }

\subsection{Swift observations} 
\label{ob_swift}

The Swift X-Ray Telescope \citep[XRT; ][]{Burrows05} data were reduced using the \swift\ 
tools within {\sc heasoft} v. 6.23 \citep{Blackburn95}.  All observations 
were reprocessed using the tool {\sc xrtpipeline}. The source light curves and 
spectra were extracted in the 0.3$-$10.0 keV band using a 20-arcsec circular 
extraction region centered on the source position. Background data were extracted 
from an annular region with an inner radius of 30 arcsec {\KK and an outer radius of} 60 arcsec.
In some {\KK photon counting} (PC) mode observations (with average count rates higher than 0.5 counts/s), 
about 2$-$4 pixels were removed from {\FB the center of the extraction region of the source} to account for the 
pile-up. The 15--50 keV Burst Alert Telescope \citep[BAT;][]{Barthelmy05} daily light 
curve was also downloaded\footnote{https://swift.gsfc.nasa.gov/results/transients/} 
and a 5-day binning was applied to it.

The Ultraviolet/Optical Telescope \citep[UVOT; ][]{Roming05} on board \swift\ operated in 
imaging mode during all observations. In each UVOT observation, the source was 
observed with at least one of the six available filters: $uvw2$, $uvm2$, $uvw1$, 
$u$, $b$, and $v$. The UVOT data were analyzed following \cite{Poole08}. 
We extracted the source photons from a circular region with a radius 
of 4 arcsec. Background data were extracted from an annular 
region with an inner radius of 15 arcsec {\KK and an outer radius of} 25 arcsec. We used the task 
{\sc uvotsource} to determine the optical/UV flux. 

\begin{figure*}
\begin{center}
    \includegraphics[width=7.0in, angle=0]{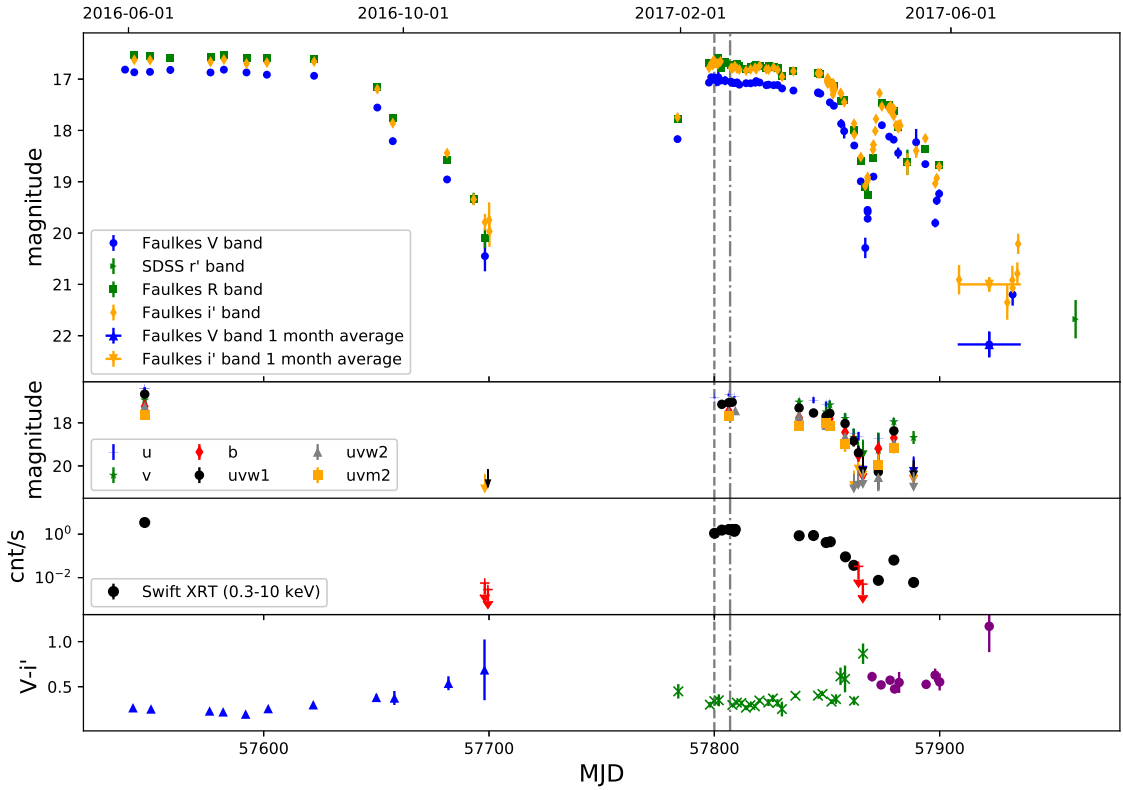}
    \caption{\textit{First panel}: Optical light curve of J1753
    during the first decay and the following mini-outburst {\KK and reflares}. The vertical 
    lines represent MJD 57800 and 57807, when the optical flux and UV/X-ray fluxes peaked, respectively. \textit{Second panel}: \swift-UVOT 
    light curve in different filters. Downward arrows show upper limits.
    \textit{Third panel}: \swift-XRT 0.3$-$10.0 keV light curve. \textit{Fouth panel}: 
    Optical color ($V$ to $i\arcmin$) of J1753 with 2-day bins. The data with triangles(blue),
    crosses(green) and filled circles(purple) were observed during the first 
    fade, the mini-outburst and {\KK the reflares}, respectively. }
    \label{fig_lightcurve_x_uv_zoom}
\end{center}
\end{figure*}

\section{Results}
\label{result}

\subsection{First rapid flux decay}

We show the optical and X-ray (15$-$50 keV) light curves of J1753 in Fig. \ref{fig_lightcurve_x_uv}.
The {\KK beginning} of the X-ray light curve shows { the decay of} a FRED type outburst
{ \citep[see also Figure 1 in][]{Zhang07}}. This is followed by 
an unusually long period of low-level activity ($\sim$11 years), 
with several lower intensity peaks. After a short transition to a
low-luminosity soft state in 2015 March--April \citep[MJD $\sim$57100;][]{Shaw16b}, 
the source flux went again below the BAT detection limit by the end of 
September 2016 (MJD $\sim$57660), when J1753 faded toward quiescence.

After the initial outburst peak in 2005, the optical light curve remained remarkably 
steady, around $V \sim 16.6$; $i\arcmin \sim 16.4$ mag from 2010 (MJD $\sim$55200) to 
2014 (MJD $\sim$57000). The optical flux then gradually faded by 0.6 mag in both $V$ 
and $i\arcmin$-bands, during the 2015 transition to the low-luminosity soft state, 
before subsequently recovering back in the hard state. From 2016 September (MJD 
$\sim$57630) onward, the optical flux started a rapid decay (hereafter, the ``first decay'') 
and within $\sim 3$ months (by MJD 57700) the optical flux had faded by a factor of $\sim 30$, 
becoming consistent with the quiescent {\KK flux upper} limit \citep[V$\sim21$ mag;][]{CadolleBel07,ATel9741}.
Assuming an exponential decay of the optical flux, we measure from MJD $\sim 57630$ to 
MJD $\sim 57700$ an average decay rate of $\sim$0.051(3), 0.059(2), and 0.052(1) mag/day 
in $V$, $R$, and $i\arcmin$-bands, respectively. All quoted errors in the text are at 
$1\sigma$ c.l. if not otherwise specified.

We show in Fig. \ref{fig_lightcurve_x_uv_zoom} a zoom-in of the multi-wavelength 
optical light curve from 2016 May onward {\FB to highlight the rebrightening episodes}.
Only three observations were taken by \swift\ during 2016; before {\KK the first decay}, and 
on November 6 and 7 near the end of the {\KK first decay} (MJD 57699). The source {\KK could not be detected} with 
XRT and UVOT in November \citep[{\KK or with radio facilities};][]{ATel9765,Plotkin17}, 
further implying that the source was close to or in 
quiescence for the first time since its discovery \citep{ATel9735}. Then, up to 
the end of January 2017, the source was Sun constrained and not visible from the ground.

\subsection{Bright mini-outburst}

When J1753 was again visible from the ground, on January 30, 2017 (MJD 57783), it became 
clear that it had unexpectedly brightened at optical wavelengths 
\citep[][see Fig. \ref{fig_lightcurve_x_uv_zoom}]{ATel10075}.
Subsequent radio \citep{ATel10110}, UV, and X-ray observations confirmed the 
flux increase in all bands \citep{ATel10081}. { According to our {\FB new} classification 
method, {\KK this} rebrightening after the main outburst is {\KK a} mini-outburst 
{\FB (i.e. the flux reached quiescence,  $\Delta t_{\rm 1} / t_{\rm out}\ll 1$, 
and $F_{1}/F_{out}\sim0.35$ in the V-band)}. The total duration of the 
{ first peak of the mini-outburst} is $\sim90$ days (MJD 57780--57870).}

We find that the optical, UV and X-ray light curves are well correlated from MJD 57800 onward. 
In the X-ray--UV bands the mini-outburst seems to peak around MJD 57807
({\KK dot-dashed} vertical line in Fig. \ref{fig_lightcurve_x_uv_zoom}), {\KK which is} about seven days after 
the optical flux peak (dashed vertical line around MJD 57800). However, there are no UV 
data taken at the time of the optical peak 
{\FB and the UV and X-ray coverage is sparse compared to that in the optical bands.}

Interestingly, we find that the optical flux peaks at a value consistent with the extrapolation 
of the slow fading before the first decay, suggesting the source somehow retains information 
about the previous flux decay \citep[this was also the case for the first mini-outburst 
of GRO J0422+32 in 1993; e.g.][]{Callanan95}. {\KK After the peak,} J1753 remained bright for approximately 50 days, slowly fading. 
%{ We used a linear function to describe the lightcurve during a certain time interval in the decay. }
During the slow fade, the optical flux first decreased up to MJD $\sim$57845 at a rate of 
0.0037(5), 0.0039(6) and 0.0034(4) mag/day in $V$, $R$ and $i\arcmin$-bands, respectively
{\FB (assuming an exponential decay)}. 
After MJD $\sim$57845, the optical flux started to decrease rapidly {\FB (assuming an exponential decay)} at a rate of 
0.158(8), 0.112(4), and 0.098(4) mag/day in $V$, $R$, and $i\arcmin$-bands 
%(UV and X-ray also decay faster)
{\KK (with a corresponding decay in UV and X-ray fluxes as well)}. 
This is about $\sim 2-3$ times faster than the {\KK decay} rate 
of the first decay in 2016.

{\KK After} reaching { a minimum flux corresponding to} $V \sim 20.3$ {\FB mag} on MJD 57866, 
the optical flux started to rise again. We recorded a new 
{ rebrightening} {\KK episode} peaking at $V=$ 17.9 {\FB mag} on MJD 57874 \citep{ATel10325}. 
{ Since the source did not reach quiescence before this second rebrightening, its properties 
have to be compared with that of the mini-outburst of which it is part of. In particular, 
{\KK $F_{2}/F_{out} \sim 0.4$, and the slope after the rebrightening peak is different from the 
slope of the mini-outburst decay,}. Consequently, we classify {\KK the second rebrightening} as a 
$\sim$45 days long reflare (of the mini-outburst). }
{\FB There is at least another reflare shortly after the first (MJD 57889).}

After the {\FB reflares the source went to quiescence again (likely between MJD 57900 and 57908)}. We combined our quiescent data into deep $V$ and $i\arcmin$-band images, 
and calculated an average quiescent magnitude of $V=$ 22.17 $\pm$ 0.25; 
$i\arcmin$ $=$ 21.00 $\pm$ 0.14 during June 2 to July 5, 2017 \citep[see also][]{ATel10562}.

\begin{figure*}
    \centering
    \includegraphics[height=6.48in, angle=270]{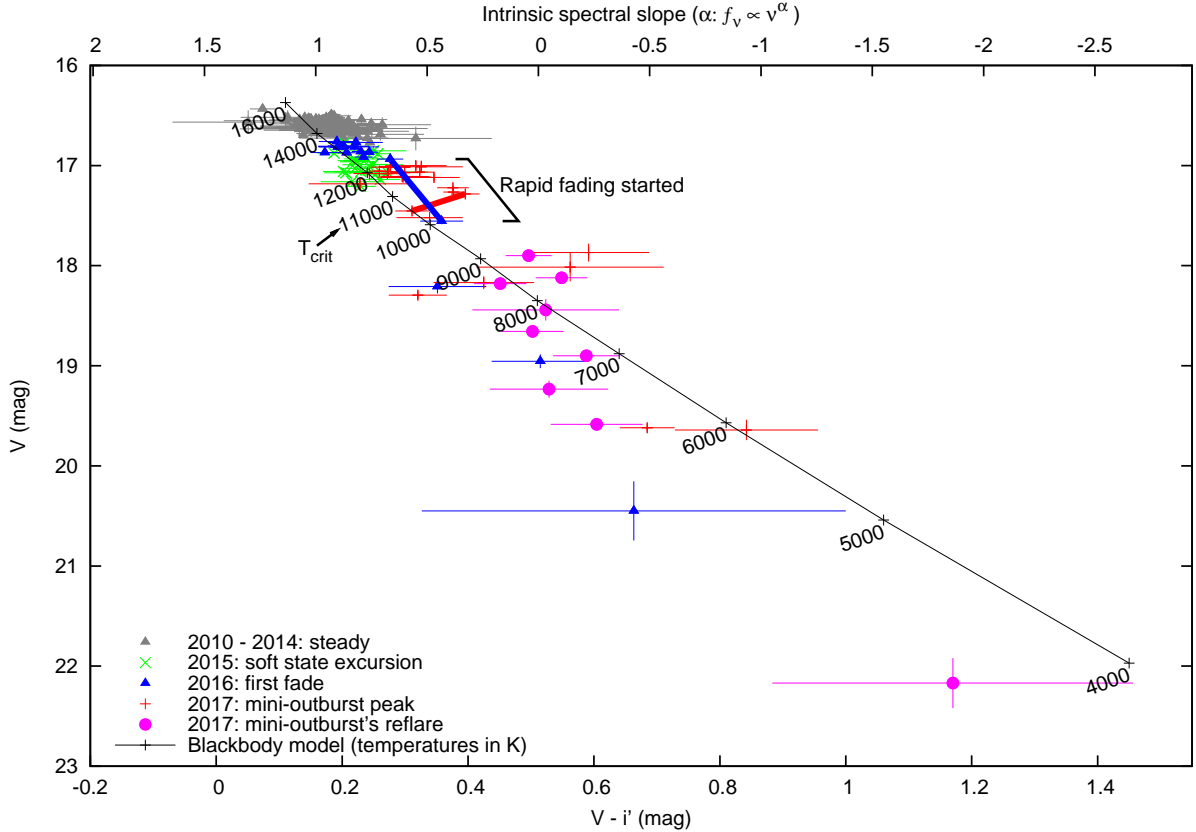}
    \caption{Colour-magnitude diagram (CMD) of the entire optical 
    light curves. Blue colors are towards the left, red colors 
    towards the right. Data were included when $V$ and $i^{\prime}$ 
    magnitudes were obtained within two days.  
    The lower axis shows the (not de-reddened) optical color between $V$ 
    and $i^{\prime}$-bands. The upper axis shows the corresponding 
    intrinsic (de-reddened) spectral index $\alpha$. The symbols denote 
    the dates when the data were taken, with the epochs of soft state 
    (2015), fading and {\KK rebrightening events} (2016--2017) shown as different 
    symbols. The black solid line shows the blackbody model (see text) 
    with temperatures in Kelvin indicated, ranging from 4000 K near 
    quiescence, to 16,000 K near outburst peak. The data generally 
    appear to follow the trend of the model, with some small excursions 
    away from (redder than) the model. For the first rapid fade in 2016 
    and the fade of the mini-outburst in 2017, the fading started 
    just as the source crossed the critical temperature $T_{\rm crit}$ 
    indicated in the figure (the blue and red thick lines indicate when 
    the source crossed $T_{\rm crit}$).
    }
    \label{fig_optical_cmd}
\end{figure*}

\subsection{Color evolution}

Throughout the fading and rebrightening periods there were large variations in the 
optical $V$ to $i\arcmin$ color (Fig. \ref{fig_lightcurve_x_uv_zoom}, {\FB bottom panel}). 
In Fig. \ref{fig_optical_cmd} we show the color--magnitude diagram (CMD), 
using 2-day bins of the data. The symbols correspond 
to different stages of the evolution of J1753. The color is bluer when brighter, 
and the general trend of the whole data set is well described by a simple model 
\citep{Maitra08,Russell11} of a single temperature, constant area blackbody heating 
and cooling. At the lower temperatures, the optical emission originates in 
the Rayleigh-Jeans tail of the blackbody, whereas at higher temperatures it is 
coming from near the curved peak of the blackbody; this causes the colour changes.
We adopt the same method as \cite{Russell11} to apply this model to 
the data of J1753. The model 
assumes 
an optical extinction of E(B-V)=$0.45\pm0.09$ \citep{Froning14}, the \cite{Cardelli89} 
extinction law and the { an orbital period of $P_{ orb}$ = 3 hr. }
The source consistently follows the model 
{\KK during the fading and brightening periods of the mini-outburst and the reflares,}
implying that the optical emission is dominated by a characteristic 
disk radius with a temperature of $\sim 15,000$ K during the period 
{\KK from} 2010 {\KK to} 2015, cooling down to $\sim 4,000$ K {\KK when the source is} near quiescence (although the companion 
star may dominate at these low fluxes). This region has to be either the thermally emitting 
outer regions of the disk, or the 
{\KK irradiated outer disk} \citep[e.g.][]{Hynes05}. We cannot 
distinguish this based on the emission area, since this depends on the uncertain distance 
of the source. However, in \cite{Shaw19}, sophisticated broadband fitting 
is applied to data from the {\KK rebrightening episodes} of J1753, and a fully irradiated disk is favoured 
from their modeling.

According to the irradiated DIM (IDIM), as the irradiated accretion disk gradually 
cools during the initial decay after outburst peak, 
a cooling wave develops in the outer disk
that is at a critical photospheric temperature, shutting down the outburst rapidly (\citealt{King1998}, D01). 
This temperature corresponds 
to the minimum column density for the upper branch of the `S-curve', which for an 
irradiated disk is $T_{\rm crit}\sim 11,000$ K \citep[see Equation A.2 in][]{Lasota08}. 
We find from Fig. \ref{fig_optical_cmd} that 
the temperature of the emitting region is $\sim 11,000-12,000$ K just when the source 
starts to fade rapidly. This occurs twice -- once at the start of the first {\FB decay} 
in 2016 (MJD 57622--57650) and again near the end of the mini-outburst in 
2017 (MJD 57848--57852). These are shown in the CMD in Fig. \ref{fig_optical_cmd} 
as thick blue and red lines, respectively. On both occasions the source crosses 
$T_{\rm crit}$, immediately fading rapidly, as expected in the 
IDIM as the cooling wave propagates through the disk. A similar process may have 
occurred in GRO J0422+32, in which there is a similar bluer-when-brighter behavior 
during the main {\FB outburst} and {\FB the following two rebrightening events, that according 
to our classification method are two mini-outbursts} \citep[see Fig. 4 of][]{Chevalier95}.

This can explain why the source did not fade rapidly before 
2016 -- because the outer disk temperature remained above $T_{\rm crit}$, so the 
entire disk was likely to be above the critical temperature. While the disk remained in this 
quasi-steady state above $T_{\rm crit}$, the mass transfer rate from the companion star must 
have been approximately equal to the mass accretion rate onto the compact object, since the 
viscous timescale is on the order of a month; much shorter than the 11-year `standstill' period. In addition, there was 
no slow fade during the second mini-outburst in May--June 
2017, because the disk remained \emph{below} $T_{\rm crit}$, only reaching 
$\sim 9,000$ K, so it was likely that cooling fronts were still active, 
rapidly ending the outburst.

We note that there appear to be excursions away (redder) from the blackbody model 
near the peaks of the two mini-outbursts in Fig. \ref{fig_optical_cmd}. This could be 
due to a changing area of the blackbody, or a separate transient component redder than 
the disk that contributed to the optical flux at the time \citep[e.g. synchrotron 
emission from the jet;][]{Russell11}.

\section{Discussion}
\label{discussion}

{\KK The characteristics of the outburst of J1753 are peculiar in many ways. 
Three aspects of the outburst evolution are particularly interesting and} not easily explained under the IDIM. 

\begin{itemize}
\item Firstly, the 11-year steady `standstill' period following the initial FRED is atypical, { and possibly unique}. 
%As mentioned above, 
This requires the mass accretion rate onto the black hole to be comparable 
to the mass transfer rate from the companion star. We find that the standstill can be due to 
the outer disk not cooling sufficiently, 
$T_{\rm crit}$ not being reached, preventing a cooling front from forming. 
Why it took 11 years for the source to reach $T_{\rm crit}$ is unknown, although the 
2015 low-luminosity soft state provides a clue as it precedes the rapid fade. J1753 is a short orbital 
period system so its irradiated disk is relatively small, so it is essentially missing the cool, outer 
regions of a larger disk. This may have played a role in keeping the whole disk hotter than the critical temperature.
\item Secondly, rather than having low-amplitude reflares during the outburst fade, 
the source {\FB first} reached close to quiescence and then {\FB suddenly} brightened by $> 3$ magnitudes 
in optical and by $\sim 3$ or more orders of magnitude in X-ray {\FB during the mini-outburst}. The amplitude of 
the rise into the mini-outburst is greater than expected from reflares, 
which have optical amplitudes of $\sim 1$--2 mag at most (D01). 
\item Thirdly, the recurrence time between the end of the outburst and the start of 
the mini-outburst is very short; only a few months or less. 
This is insufficient time for the disk to be evaporated and filled in {\FB again} before the mini-outburst.
A further clue regarding the timing of the mini-outburst was suggested by \cite{Plotkin17},
see also \cite{Shaw13}. 
They identified a long-term X-ray modulation during the standstill, with a period of $\sim 400$ days 
and pointed out that the mini-outburst might be an extension of this long-term modulation (see their Fig. 3).

\end{itemize}

Mini-outbursts are rare {\KK events}
whereas reflares are commonly reported 
in the optical and X-ray light curves of BHXTs.
Similar mini-outbursts have been observed from the BHXTs GRO J0422$+$32 
\citep[R-band,][]{Shrader94,Chen97}, MAXI J1659--152 \citep[X-rays,][]{Homan13}, 
GRS 1739--278 \citep[X-rays,][]{Yan17,CorralSantana18}, A0620--00 \citep[optical,][]{Charles98}
{\FB and possibly GRS 1009$-$45 \citep[V-band,][]{Chen97}}.

The above systems all have short orbital periods, 
in the range $P_{\it orb}\sim$2.5--7 h except GRS 1739--278, for which $P_{\it orb}$ 
is unknown, and we here suggest it may have a short {\FB orbital} period. 
In fact, mini-outbursts have been detected from three out of the five BHXTs with periods $\leq 5.1$ h \citep[see table 2 in][]{Shahbaz13}. The other two are XTE J1118$+$480 \citep{Chaty03}, 
which in 2000 had a { multi-peak} outburst {\FB according to our classification}, 
and Swift J1357.2--0933, which did 
not have a mini-outburst following its 2011 or 2016 outbursts \citep[but coverage was poor 
in the months after the 2011 outburst; see][ and references therein]{Russell18}.
The optical, X-ray, and radio \citep[see also][]{Plotkin17,Shaw19}
monitoring of the {\FB mini-outburst and following reflares of J1753} represent the most complete 
multi-wavelength coverage of {\FB such an event} for a BHXT to date.

The optical light curves of the BHXT mini-outbursts are morphologically very similar 
to those of another class of short orbital period (few hours) binaries, the dwarf novae 
(DN) of the RZ LMi-type \citep[][and references therein]{Hameury00}
which exhibit short `superoutburst' duty cycles of as low as $\sim 20$ days 
\citep[e.g.][]{Osaki95}. Dwarf novae are cataclysmic variables, i.e. close binary 
systems that host accreting white dwarf 
primaries, and show much shorter outburst durations and recurrence times 
than BH or NS X-ray transients \citep[see][]{Schreiber03,Britt15,Zhang17c}. Intriguingly, the first mini-outburst following a normal outburst of a BHXT or DN seems to usually peak at a flux level consistent 
with the extrapolation of the decay curve that precedes it. 
Mini-outbursts may be closely related to a critical temperature of the accretion disk 
corresponding to that flux. For J1753 we have demonstrated that this is indeed 
the case; when the rapid fade begins, the outer disk 
temperature is cooling and crossing the critical photospheric temperature needed 
for a cooling wave to develop, shutting down the outburst.

The DIM is able to reproduce the outburst light curves of DN mini-outbursts
under the condition that the inner disk remains hot at the end of an outburst. 
In this case, when a cooling front develops in the disk the density/temperature 
immediately below the front
is high enough to start a new heating wave. 
The inward moving cooling front is `reflected' and an outgoing 
heating front starts a new (mini)outburst (see e.g. D01).

Perhaps the same mechanism could cause mini-outbursts in BHXTs if, like in DN, 
there is a hot inner disk near the 
end of the outburst. There were no soft X-ray observations during the first 
fade of J1753 in 2016, so the inner disk radius is unknown at this time. However, 
in 2015 the source was observed in a { unusual } faint soft state, during which \cite{Shaw16b} measured 
a disk inner radius of R$_{\rm in} \sim 12$--$28 R_{g}$, at a temperature of 
$kT_{\rm in} = 0.25$ keV and an X-ray luminosity 
of $\sim 6 \times 10^{-3} L_{\rm Edd}$. J1753 is therefore certainly capable of 
maintaining a hot inner disk at low luminosities. Presumably, during the first 
fade of J1753 in 2016 the source was in the low-hard state, which is 
usually characterized by a truncated disk. However, in the low-hard state of 
J1753 before 2016, evidence for the presence of a faint inner disk that resides at 
$< 20 R_{g}$ has been provided by iron line fitting, a soft excess in the 
continuum and variability spectra \citep[][]{Wilkinson09, Kalamkar13, 
Tomsick15, Kajava16}. This is unusual for a BHXT in the low-hard state, 
and this could be an additional reason for J1753 to have { a mini-outburst} while 
many other (longer period) systems do not.

We propose that BHXT mini-outbursts are actually fairly common in short-period 
{\FB ($< 7$ h)} BHXTs, and may be due to the presence of a hot inner disk at the 
end of the outburst decay, and arise from a sequence of heating and cooling 
front reflections in the  accretion disk. Testing this would require a deep 
soft X-ray observation during an outburst fade, to confirm the presence of a 
hot inner disk just before a mini-outburst. We speculate that in BHXTs with 
longer orbital periods, their disks are larger and therefore the outermost 
regions of their disks are cooler than in short period BHXTs.  The inward 
traveling cooling front may therefore take longer to reach the hot inner 
regions of the disk in long period systems compared to short period systems, 
by which time the inner disk may be truncated. Truncation typically occurs over 
several days--weeks during an outburst decay, which is a similar order of 
magnitude to the timescale for a heating/cooling front to move in/out 
\citep[see e.g.][]{Bernardini16}.
In the short period systems, the temperature at the outer radius of their 
smaller disks will be higher than the temperature at the outer radius of the 
larger accretion disks of the longer period systems (for the same accretion rate). 
The lack of the cool outer regions in short period XTs may cause the 
heating/cooling fronts to remain hot enough to initiate a new (mini)outburst, 
or the inner disk truncation radius may be somehow regulated by the small 
size of the disk and/or the hot flow at smaller radii.
In addition, the hard-state nature of most of the outburst of J1753 
(and most other short period BHXTs) provides a clue about the state of the disk, 
i.e. that the irradiating source is hard and possibly extended (the hot flow / Comptonizing 
region is likely to be physically thicker than the inner disk), making it able to illuminate 
and heat the outer disk more readily than the soft photons from the inner disk. 
It is worth noting that long outbursts lasting several years or decades have 
been reported in some long period XTs, e.g. GRS 1915+105 ($>20$ years; 
$P_{\it orb} = 33.5$ days), GRO J1655--40 (16 months; $P_{\it orb} = 2.6$ days), 
KS 1731--260 (12.5 years; $P_{\it orb} \sim 10$ hours) \citep*{Sobczak99,Wijnands01,Deegan09,Zurita10}, 
although V404 Cyg had two very short outbursts in 2015--2016 \citep{MunozDarias17}. 
However, the neutron star system EXO 0748--676 had an outburst lasting 24 years, 
and it has a 3.8 hr period \citep[e.g.][]{HynesJones09}.

%######################################333

\subsection{Standstill}  
Since the beginning of the J1753 outburst is so well explained by the IDIM, it is 
worth investigating what makes its end different from what the standard version of 
the model predicts. During the long standstill period, the outer disk temperature is observed to 
be $\sim 15,000$ K. This is most likely an irradiation 
temperature on the disk surface\footnote{$F_{\rm irr}/F_{\rm vis} \sim {\mathcal  C}R/R_{\rm in}$, 
so for $R/R_{\rm in} \gtrsim 10^3$, the irradiation flux is always larger 
than the flux produced by viscosity.}. Therefore
\begin{equation}
T_{\rm irr}= \left({\mathcal  C}\frac{\dot Mc^2}{4\pi \sigma_{\rm SB}R^2}\right)^{1/4}\approx 18,800 \, {\mathcal  C}^{1/4}_{-3}{\dot M_{16}}^{1/4}R_{10}^{-1/2}\,\rm K,
\label{Tirr}
\end{equation}
where $\dot M=\dot M_{16}10^{16}\rm\,g\,s^{-1}$. The mean flux during standstill 
is $F_{X}^s\approx  10^{-9}\,\rm erg\,cm^{-2}s^{-1}$ corresponding to an 
accretion rate $\dot M_{s}\approx 1.3 \times 10^{16}\rm g\,s^{-1}$.
For such an accretion rate, a temperature of $\sim 15,000$ K, corresponds to 
$R\approx 1.8\times 10^{10}\, {\mathcal  C}^{1/2}_{-3}$cm.

On the other hand, since the standstill corresponds to a steady stable disk, one 
must have  $\dot M_{s} > \dot{M}_{\rm crit}^{+}(R_D^s)$, where $R_D^s$ is the disk outer 
radius during standstill. The disk radius shrinks during the outburst decay, and a 
sudden enhancement of the mass-transfer rate might reduce its 
value. For $\dot M_{s}= 1.3 \times 10^{16}\rm g\,s^{-1}$, the disk during standstill 
is stable when $R_D^s < 4.8\times 10^{10}{\mathcal  C}^{0.15}_{-3}$cm, consistent 
with the value deduced from the irradiation temperature. A fluctuation lowering the mass transfer rate from the companion star could end the standstill and trigger the decay of the outburst. Note that a detailed 
analysis of the standstill period based on broadband SED modelling is presented 
in \cite{Shaw19}.

\subsection{The outburst of J1753 according to the IDIM}
Here, we investigate how the initial decay from maximum of the J1753 outburst can be 
described by the Irradiated DIM (IDIM). We assume a distance $d= 3$ kpc, 
a black-hole mass in solar 
units, $m=7$, and an orbital period 
$P_{\rm orb}=3$ hr. Using for the maximum disk radius the formula \citep[see e.g.][]{Lasota08}
\begin{equation}
R_{D}({\rm max}) = 2.1\times 10^{10}m^{1/3}P_{\rm orb}^{2/3}\rm cm,
\label{radmax}
\end{equation}
where $P_{\rm h}$ the orbital period in hours, one obtains for J1753 $R_{D}({\rm max})=8.3\times 10^{10}$cm.

According to the IDIM, the maximum accretion rate (at the luminosity peak) is 
that of a (quasi)steady  disk accreting at constant rate of $\sim 3\dot{M}_{\rm crit}^{+}(R_D)$, where
\begin{equation}
\dot{M}_{\rm crit} \approx 9.5 \times 10^{14}~{\mathcal  C}_{-3}^{-0.36}R_{D,10}^{2.39}~m^{-0.64}\,\rm g\,s^{-1},
\label{Mdotcrit}
\end{equation}
is the value of the minimum critical accretion for a hot, irradiated disk at 
its outer radius $R_{D}=R_{D,10}10^{10}$cm \citep{Lasota08}, and ${\mathcal  C}=10^{-3}{\mathcal  C}_{-3}$ 
is a constant characterizing the outer-disk irradiation by a central point-like 
source. We ignore the very weak dependence on the viscosity parameter $\alpha$.

Assuming  ${\mathcal  C}_{-3}=1$ (\citealt{Lasota08}; see however \citealt{Tetarenko2018}), one obtains for the accretion 
rate at maximum of the J1753 outburst $\dot M_{\rm max}\approx 1.3 \times 10^{17}\rm g\,s^{-1}$. 
For an accretion efficiency $\eta=0.1$ this corresponds to a peak luminosity 
$L_{\rm max}\approx 1.2 \times 10^{37}\,\rm erg\,s^{-1}$. The observed X-ray 
flux at the peak of J1753 outburst was $F_{X}^p\approx  10^{-8}\,\rm erg\,cm^{-2}s^{-1}$ 
which at the assumed distance corresponds to  $L_{\rm max}\approx 1.2 \times 10^{37}\,\rm erg\,s^{-1}$, which is 
in excellent agreement with the IDIM. Obviously this excellence is a coincidence 
in view of the uncertainties in the values of the distance \citep[which could well be further than the assumed 3 kpc; the GAIA parallax distance is unreliable;]{Gandhi18},  ${\mathcal  C}$ and 
$\eta$. Nevertheless, it shows that the IDIM can be successfully fit to describe the initial peak of the outburst of 
J1753, at least until the standstill.

This is confirmed by estimating the outburst decay time which according to the IDIM is
\begin{equation}
t_{\rm dec} \simeq \frac{R^2}{3\nu},
\label{tvis}
\end{equation}
where $\nu = \alpha c_s^2/\Omega$ is the Shakura--Sunyaev kinematic viscosity coefficient, $c_s \propto T_c^{1/2}$ the sound speed, $T_c$ the disk midplane temperature, and
$\Omega = (GM/R^3)^{1/2}$.
Taking the critical midplane temperature $T_{\rm c}^{+}\approx14,500\,\rm K$ corresponding to $\dot{M}_{\rm crit}^{+}$ \citep{Lasota08} and the parameters assumed for J1753, one obtains
for the decay timescale
$t_{\rm dec}\approx 93\, \alpha_{0.2}^{-1} \,\rm days$,
where $\alpha_{0.2}=\alpha/0.2$ (i.e. we adopt $\alpha = 0.2$).  This is again in excellent agreement with observations of the {\KK first} decay lasting $\sim 3$ months.

\subsection{Optical fade during the 2015 soft state}
As mentioned in Section 3.1, the optical emission appears to gradually decrease by 
$\sim 0.6$ mag in 2015 around MJD 57100, which is when the source made a transition 
to the faint soft state \citep{Shaw16b}. The optical flux then recovers, brightening 
back in the hard state, before starting the slow decay (Fig. \ref{fig_lightcurve_x_uv}). 
A fade of the optical/infrared flux is common in the soft state in BHXTs, and 
is usually attributed to the disappearance of the synchrotron jet component 
\citep[e.g.][]{Kalemci13}. However, this 
optical fade is usually more prominent in the infrared, and less at optical 
wavelengths, with a large change in the color in the soft state. Such a color 
change is not apparent in J1753; in Fig. \ref{fig_optical_cmd} we find that the 
soft state data (green crosses) lie on the blackbody line, as do the hard state 
data just before and after the soft state, with a very similar color. The optical 
fade in the soft state is therefore unlikely to be due to a synchrotron component 
disappearing in this state. An excess has been detected in J1753 in the 
hard state, but only at longer wavelengths \citep{Froning14,Rahoui15}, not at 
optical wavelengths.

The soft X-ray flux increases in the soft state, as the inner disk is close to 
the BH and the inner disk temperature increases. Concurrently the hard X-rays 
decrease, as the power law component fades in the soft state (this is visible 
in the Swift BAT light curve; Fig.\ref{fig_lightcurve_x_uv}). If the optical 
emission originates in the outer regions of the disk, the optical flux should 
remain approximately constant over the transition to/from the soft state 
\citep[e.g.][]{Russell11}. If the {\KK irradiated outer accretion disk} produces the optical 
emission instead, the optical will be positively correlated with the X-ray 
emission. The optical flux of J1753 does not react to the increase of soft 
X-rays, which implies it would be reprocessed hard X-rays. Since the X-ray 
power law in the hard state likely resides up to at least $\sim 100$ keV 
\citep[e.g.][]{Tomsick15,Kajava16}, and may even extend beyond 600 keV 
\citep{CadolleBel07}, the bolometric luminosity is likely to be higher 
in the hard state than in the soft state. Moreover, as well as being more energetic, 
harder X-rays can penetrate further into the disk atmosphere, and so heat the disk more efficiently. In addition, if the X-ray power 
law is elevated above the disk, such as in the lamppost model for the 
Comptonized region, these hard photons will more readily illuminate the outer disk, compared to the 
inner regions of the disk itself. To summarize, the optical fade in the 
soft state is either caused by a decrease of intrinsic emission from the 
outer disk (for some unknown reason), or a decrease in the irradiation bump 
(if this is reprocessed hard X-rays).

\section{Conclusions}
\label{conclusions}
 
In this paper we have presented new optical, UV and X-ray observations of the short orbital period 
BHXT \src, focusing on the final stages of its 12$-$year long outburst that started in 2005.

(i) We recorded a bright mini-outburst { with a  reflare} at the end of the main outburst decay. 
We found from the
optical colors that the temperature of the outer disk was $\sim 11$,000 K when the source started 
to fade rapidly on both occasions. The  mini-outburst had a peak flux consistent with the 
extrapolation of the previous decay before the first fade, which is similar to some mini-outbursts 
in other BHXTs.

(ii) According to the IDIM, the $\sim 11$,000 K is consistent with being the critical temperature (for an 
irradiated disk) when a cooling wave forms, accelerating the fade of the outburst, as is observed.  
The optical color could be a useful tool to predict decay rates in some X-ray transients. 

(iii) Based on available observations, we suggest the  X-ray binaries that have known 
mini-outbursts following an outburst are all short orbital period systems ($< 7$ h). 
Their smaller accretion disks may be a requirement for mini-outbursts to occur. Another requirement 
may be the existence of a hot inner disk at the end of the outburst, as was previously used 
to explain the mini-outbursts from RZ LMi type cataclysmic variables.

{ 
(iv) In this paper we also introduced {\FB for the first time} a new method to classify rebrightening {\KK events} quantitatively 
in NS/BH-XRBs. 
}

\section*{Acknowledgments}
The Faulkes Telescopes are maintained and operated by the Las Cumbres Observatory (LCO).
The X-ray data are obtained from the High Energy Astrophysics 
Science Archive Research Center (HEASARC), provided by NASA's Goddard
Space Flight Center and NASA's Astrophysics Data System Bibliographic Services. 
We thank the \swift\ Science Operations Team for performing the observations.
The research reported in this publication was supported by Mohammed Bin Rashid Space 
Centre (MBRSC), Dubai, UAE, under Grant ID number  201701.SS.NYUAD.
GZ acknowledges funding support from the CAS Pioneer Hundred Talent Program Y7CZ181001.
FB is founded by the European Union's Horizon 2020 research and innovation programme 
under the Marie Sklodowska-Curie grant agreement no. 664931.
AWS is supported by an NSERC Discovery Grant and a Discovery Accelerator Supplement. PAC is grateful 
to the Leverhulme Trust for the award of a Leverhulme Emeritus Fellowship.
RMP acknowledges support from Curtin University through the Peter Curran Memorial Fellowship.
LCO data from this work are being used by FL and RD in their educational project, 
``Black Holes In My School''. PK (teacher) and DP (pupil) represent one of the pilot schools in this project. 
JPL was supported by the Polish NCN grant 2015/19/B/ST9/01099 
and the French National Space Center CNES. JCAMJ is the recipient of an Australian Research 
Council Future Fellowship  (FT140101082).
\section{appendix}

\begin{table*}
\begin{center}
\caption{Log of LCOGT observations from June 2016. }
\begin{tabular}{ccccccccc}
\hline\hline
MJD &  magnitude($i\arcmin$) & MJD & magnitude($i\arcmin$) & MJD & magnitude($R$) & MJD & magnitude($V$) \\
\hline
 57576.408966 & 16.664$\pm$0.041 &  57852.653244 & 17.206$\pm$0.007 &  57558.574655 & 16.858$\pm$0.040 &  57558.571983 & 16.821$\pm$0.075 \\
 57582.337375 & 16.619$\pm$0.033 &  57852.667005 & 17.147$\pm$0.008 &  57576.406211 & 16.849$\pm$0.007 &  57576.403534 & 16.871$\pm$0.012 \\
 57592.379523 & 16.698$\pm$0.029 &  57852.673729 & 17.136$\pm$0.013 &  57582.334629 & 16.792$\pm$0.008 &   57582.33196 & 16.816$\pm$0.014 \\
 57601.423749 & 16.680$\pm$0.024 &  57852.678244 & 17.135$\pm$0.008 &  57592.376769 & 16.853$\pm$0.012 &  57592.374098 & 16.870$\pm$0.019 \\
 57622.373905 & 16.659$\pm$0.033 &  57852.686403 & 17.139$\pm$0.009 &  57601.420983 & 16.868$\pm$0.008 &  57601.418325 & 16.914$\pm$0.012 \\
 57650.438382 & 17.196$\pm$0.030 &  57852.691101 & 17.118$\pm$0.013 &  57622.371154 & 16.889$\pm$0.010 &  57622.368481 & 16.936$\pm$0.019 \\
 57657.303600 & 17.858$\pm$0.031 &  57852.698035 & 17.115$\pm$0.037 &  57650.435627 & 17.427$\pm$0.012 &  57650.432968 & 17.554$\pm$0.031 \\
 57681.395282 & 18.441$\pm$0.024 &  57852.709628 & 17.079$\pm$0.035 &  57657.300838 & 18.029$\pm$0.027 &  57657.298173 & 18.209$\pm$0.068 \\
 57693.199752 & 19.344$\pm$0.024 &  57852.717024 & 17.157$\pm$0.114 &  57681.392516 & 18.845$\pm$0.036 &  57681.389858 & 18.956$\pm$0.069 \\
 57698.204232 & 19.786$\pm$0.022 &   57852.72825 & 17.114$\pm$0.161 &  57693.196992 & 19.603$\pm$0.117 &   57698.19881 & 20.450$\pm$0.296 \\
 57700.039837 & 19.747$\pm$0.026 &  57852.729595 & 17.115$\pm$0.346 &  57698.201471 & 20.371$\pm$0.189 &  57783.663806 & 18.169$\pm$0.064 \\
 57700.194786 & 19.967$\pm$0.020 &  57852.734499 & 17.108$\pm$0.300 &  57783.666481 & 18.047$\pm$0.033 &  57797.623031 & 17.068$\pm$0.046 \\
 57783.669241 & 17.744$\pm$0.020 &   57852.74075 & 17.089$\pm$0.047 &   57797.62571 & 16.953$\pm$0.018 &  57798.621437 & 16.965$\pm$0.038 \\
 57797.628466 & 16.779$\pm$0.019 &  57852.746995 & 17.102$\pm$0.015 &  57798.622952 & 16.982$\pm$0.022 &  57799.386816 & 17.010$\pm$0.031 \\
 57798.619832 & 16.700$\pm$0.027 &  57852.753815 & 17.132$\pm$0.016 &   57799.38975 & 16.963$\pm$0.018 &  57799.775908 & 16.967$\pm$0.066 \\
 57799.383835 & 16.728$\pm$0.025 &  57852.762352 & 17.184$\pm$0.018 &  57799.778842 & 16.956$\pm$0.041 &  57800.773383 & 17.024$\pm$0.116 \\
 57799.772894 & 16.627$\pm$0.020 &  57852.765903 & 17.157$\pm$0.035 &  57800.774915 & 16.935$\pm$0.041 &  57801.384364 & 17.042$\pm$0.032 \\
 57800.771778 & 16.695$\pm$0.019 &  57852.773000 & 17.100$\pm$0.033 &  57801.387291 & 16.943$\pm$0.020 &  57801.769249 & 16.965$\pm$0.169 \\
 57801.381370 & 16.706$\pm$0.020 &  57852.778253 & 17.116$\pm$0.023 &  57801.770752 & 16.855$\pm$0.051 &  57802.768061 & 17.028$\pm$0.068 \\
 57801.767629 & 16.695$\pm$0.021 &  57853.033500 & 17.161$\pm$0.041 &  57801.773653 & 16.942$\pm$0.032 &  57804.766226 & 17.033$\pm$0.085 \\
 57802.765060 & 16.657$\pm$0.075 &  57853.475840 & 17.202$\pm$0.051 &  57802.770988 & 17.047$\pm$0.054 &  57804.772667 & 17.028$\pm$0.082 \\
 57807.466797 & 16.782$\pm$0.062 &  57856.258205 & 17.288$\pm$0.017 &  57804.768896 & 16.947$\pm$0.038 &  57807.394323 & 17.051$\pm$0.019 \\
 57808.111048 & 16.793$\pm$0.031 &  57856.050972 & 17.269$\pm$0.015 &  57804.775336 & 16.951$\pm$0.042 &  57807.463757 & 17.057$\pm$0.019 \\
 57809.108328 & 16.746$\pm$0.043 &  57857.626176 & 17.452$\pm$0.017 &   57807.46984 & 16.987$\pm$0.013 &  57808.108029 & 17.075$\pm$0.018 \\
 57810.105714 & 16.798$\pm$0.035 &  57861.993962 & 17.869$\pm$0.020 &  57808.114082 & 17.031$\pm$0.012 &  57809.105297 & 17.078$\pm$0.020 \\
 57811.102868 & 16.813$\pm$0.038 &  57862.094179 & 18.079$\pm$0.017 &  57809.111372 & 16.997$\pm$0.013 &  57810.102676 & 17.064$\pm$0.029 \\
 57814.101241 & 16.837$\pm$0.039 &  57865.000261 & 18.517$\pm$0.017 &  57810.108736 & 16.969$\pm$0.015 &  57811.099824 & 17.108$\pm$0.021 \\
 57816.086183 & 16.810$\pm$0.092 &  57866.986331 & 19.081$\pm$0.014 &   57811.10591 & 17.022$\pm$0.014 &  57814.104309 & 17.080$\pm$0.019 \\
 57818.080712 & 16.779$\pm$0.053 &  57868.000318 & 18.976$\pm$0.018 &  57814.107262 & 17.065$\pm$0.013 &  57816.089192 & 17.082$\pm$0.017 \\
 57818.570966 & 16.812$\pm$0.041 &   57868.04205 & 18.920$\pm$0.008 &   57816.09213 & 17.035$\pm$0.011 &  57818.083761 & 17.070$\pm$0.020 \\
 57820.075244 & 16.743$\pm$0.041 &  57868.050433 & 18.908$\pm$0.015 &   57818.08672 & 17.014$\pm$0.013 &  57818.565527 & 17.039$\pm$0.016 \\
 57823.067088 & 16.808$\pm$0.033 &  57870.405355 & 18.377$\pm$0.020 &  57818.568201 & 16.995$\pm$0.009 &  57820.078283 & 17.066$\pm$0.018 \\
 57824.064272 & 16.816$\pm$0.035 &  57870.609638 & 18.283$\pm$0.039 &  57820.081237 & 17.018$\pm$0.013 &  57823.070125 & 17.118$\pm$0.034 \\
 57826.310092 & 16.773$\pm$0.048 &  57870.625248 & 18.276$\pm$0.020 &   57823.07308 & 17.051$\pm$0.017 &  57824.067304 & 17.109$\pm$0.057 \\
 57828.053927 & 16.818$\pm$0.036 &  57871.426129 & 18.011$\pm$0.023 &  57824.070279 & 17.025$\pm$0.030 &   57826.31414 & 17.119$\pm$0.035 \\
 57830.062119 & 16.954$\pm$0.020 &  57871.752167 & 17.778$\pm$0.049 &  57826.317063 & 17.042$\pm$0.020 &  57828.056969 & 17.115$\pm$0.034 \\
 57835.098306 & 16.847$\pm$0.015 &  57873.325745 & 17.273$\pm$0.014 &  57828.059935 & 17.047$\pm$0.020 &  57830.065173 & 17.182$\pm$0.065 \\
 57846.035396 & 16.882$\pm$0.028 &  57874.361536 & 17.533$\pm$0.014 &  57830.068147 & 17.208$\pm$0.038 &  57835.101317 & 17.223$\pm$0.021 \\
 57846.045207 & 16.895$\pm$0.028 &  57877.257184 & 17.567$\pm$0.014 &  57835.104243 & 17.105$\pm$0.012 &  57846.038429 & 17.259$\pm$0.018 \\
 57847.032727 & 16.888$\pm$0.021 &  57877.615258 & 17.565$\pm$0.014 &  57846.041383 & 17.162$\pm$0.011 &  57846.048233 & 17.272$\pm$0.017 \\
 57850.244833 & 17.021$\pm$0.034 &  57878.212070 & 17.516$\pm$0.017 &  57847.038727 & 17.163$\pm$0.011 &  57847.035765 & 17.284$\pm$0.018 \\
 57850.248495 & 17.056$\pm$0.037 &  57878.587685 & 17.641$\pm$0.017 &  57851.281824 & 17.340$\pm$0.016 &    57851.2789 & 17.453$\pm$0.026 \\
 57850.257411 & 17.055$\pm$0.052 &  57879.584894 & 17.718$\pm$0.017 &  57853.039507 & 17.412$\pm$0.019 &  57853.036542 & 17.520$\pm$0.035 \\
 57850.263663 & 17.034$\pm$0.018 &  57879.572086 & 17.572$\pm$0.016 &  57856.264113 & 17.695$\pm$0.052 &  57856.261181 & 17.870$\pm$0.090 \\
 57850.270962 & 17.030$\pm$0.052 &  57880.582335 & 17.897$\pm$0.016 &  57857.623409 & 17.676$\pm$0.054 &  57857.620736 & 18.014$\pm$0.142 \\
 57850.277968 & 17.021$\pm$0.057 &  57881.579571 & 17.892$\pm$0.015 &  57862.100176 & 18.264$\pm$0.028 &  57862.097209 & 18.294$\pm$0.038 \\
 57850.284849 & 17.034$\pm$0.039 &  57881.585786 & 17.958$\pm$0.015 &  57865.006267 & 18.865$\pm$0.032 &    57865.0033 & 18.993$\pm$0.044 \\
 57850.292388 & 16.979$\pm$0.118 &  57882.580111 & 17.904$\pm$0.015 &  57866.992334 & 19.371$\pm$0.061 &  57866.989368 & 20.290$\pm$0.199 \\
 57850.343168 & 16.995$\pm$0.212 &  57885.679478 & 18.657$\pm$0.026 &  57868.006302 & 19.521$\pm$0.051 &  57868.003333 & 19.549$\pm$0.067 \\
 57850.347353 & 17.039$\pm$0.138 &  57889.557676 & 18.396$\pm$0.025 &  57870.615676 & 18.813$\pm$0.034 &  57868.046237 & 19.721$\pm$0.060 \\
 57850.354417 & 17.060$\pm$0.023 &  57893.644986 & 18.154$\pm$0.013 &  57874.365153 & 17.730$\pm$0.021 &  57868.054629 & 19.588$\pm$0.053 \\
 57850.361243 & 17.046$\pm$0.070 &  57898.038157 & 19.034$\pm$0.013 &  57877.618386 & 17.769$\pm$0.021 &  57870.612631 & 18.900$\pm$0.048 \\
 57850.366324 & 17.065$\pm$0.048 &  57898.702179 & 18.929$\pm$0.013 &  57879.575198 & 17.885$\pm$0.019 &  57874.363361 & 17.899$\pm$0.033 \\
 57850.370319 & 17.089$\pm$0.042 &  57899.770325 & 18.705$\pm$0.013 &  -- & -- &  57877.616871 & 18.121$\pm$0.038 \\
 57850.382741 & 17.007$\pm$0.286 &  57908.607852 & 20.909$\pm$0.022 &  -- & -- &  57879.573685 & 18.180$\pm$0.033 \\
 57850.384097 & 17.018$\pm$0.337 &  57930.011298 & 21.352$\pm$0.013 &  -- & -- &  57881.580338 & 18.441$\pm$0.107 \\
 57851.275926 & 17.065$\pm$0.278 &  57932.261395 & 20.916$\pm$0.018 &  -- & -- &   57889.55952 & 18.230$\pm$0.259 \\
 57852.628351 & 17.189$\pm$0.195 &  57932.394386 & 21.069$\pm$0.058 &  -- & -- &  57893.639086 & 18.656$\pm$0.044 \\
 57852.628407 & 17.191$\pm$0.220 &  57934.480869 & 20.791$\pm$0.075 &  -- & -- &  57898.041195 & 19.803$\pm$0.085 \\
 57852.637699 & 17.152$\pm$0.197 &  57934.841549 & 20.210$\pm$0.037 &  -- & -- &  57898.704951 & 19.368$\pm$0.083 \\
 57852.640745 & 17.216$\pm$0.140 &    -- & -- &  -- & -- &  57899.764824 & 19.234$\pm$0.084 \\
 57852.646999 & 17.299$\pm$1.000 &  -- & -- &  -- & -- &  57932.398345 & 21.198$\pm$0.216 \\ 
 \hline
\end{tabular}
%\caption{}
\label{tab:obs}
\end{center}
\end{table*}

\vfill\eject
\end{document}